\documentclass[aip,
    graphicx,
    amsmath,
    amssymb,
    reprint,
    floatfix,
    ]{revtex4-1}

\draft % marks overfull lines with a black rule on the right

\usepackage{graphicx} % Include figure files
\usepackage{physics}
\usepackage{caption}
\usepackage{subcaption}
\usepackage{qcircuit}
\usepackage{xcolor}

\newcommand{\LNN}{5\%}
\newcommand{\RNN}{2\%}

\newcommand{\zz}{\mathcal{ZZ}}
\newcommand{\Yb}{\textsuperscript{171}Yb\textsuperscript{+}}

\usepackage{verbatim}

\begin{document}

\setlength{\abovecaptionskip}{5pt plus 2pt minus 2pt}
\setlength{\textfloatsep}{5pt plus 2pt minus 2pt}

%\preprint{}

%TC:ignore
\title{First-Order Crosstalk Mitigation in Parallel Quantum Gates Driven With Multi-Photon Transitions \vspace{0.2cm}} %Title of paper



\author{Matthew N. H. Chow}
\email[]{mnchow@sandia.gov}
%\thanks{}
\altaffiliation{Center for Quantum Information and Control (CQuIC)}
\altaffiliation{University of New Mexico}
\affiliation{Sandia National Laboratories}

\author{Christopher G. Yale}
\affiliation{Sandia National Laboratories}

\author{Ashlyn D. Burch}
\affiliation{Sandia National Laboratories}

\author{Megan Ivory}
\affiliation{Sandia National Laboratories}

\author{Daniel S. Lobser}
\affiliation{Sandia National Laboratories}

\author{Melissa C. Revelle}
\affiliation{Sandia National Laboratories}

\author{Susan M. Clark}
\altaffiliation{Center for Quantum Information and Control (CQuIC)}
\altaffiliation{University of New Mexico}
\affiliation{Sandia National Laboratories}


\date{\today}

\begin{abstract}
We demonstrate an order of magnitude reduction in the sensitivity to optical crosstalk for neighboring trapped-ion qubits during simultaneous single-qubit gates driven with individual addressing beams. Gates are implemented via two-photon Raman transitions, where crosstalk is mitigated by offsetting the drive frequencies for each qubit to avoid first-order crosstalk effects from inter-beam two-photon resonance.
% on our trapped-ion quantum processor, the Quantum Scientific Computing Open User Testbed (QSCOUT). 
The technique is simple to implement, and we find that phase-dependent crosstalk due to optical interference is reduced on the most impacted neighbor from a maximal fractional rotation error of $0.185(4)$ without crosstalk mitigation to $\leq0.006$ with the mitigation strategy. 
%This technique is simple to implement, and when we use it on the Quantum Scientific Computing Open User Testbed (QSCOUT), our trapped ion quantum processor,
%%{\color{blue} the most spectator qubits experience significantly reduced} rotation errors.
%% the most spectator qubits experience greater than a factor of 30 reduction in observed Rabi rate due to residual illumination from neighboring beams.
%the maximum crosstalk-induced fractional change in rotation angle of the most impacted neighbor qubit is improved from $0.185(4)$ without crosstalk mitigation to $0.000(6)$ during arbitrary-phase gates run in parallel.
Further, we characterize first-order crosstalk in the two-qubit gate and avoid the resulting rotation errors for the arbitrary-axis Mølmer-Sørensen gate via a phase-agnostic composite gate.
Finally, we demonstrate holistic system performance by constructing a composite CNOT gate using the improved single-qubit gates and phase-agnostic two-qubit gate. 
This work is done on the Quantum Scientific Computing Open User Testbed (QSCOUT); however, 
our methods are widely applicable for individual-addressing Raman gates and impose no significant overhead, enabling immediate improvement for quantum processors that incorporate this technique.

\end{abstract}

% \pacs{} % insert suggested PACS numbers in braces on next line

\maketitle %\maketitle must follow title, authors, abstract and \pacs
%TC:endignore

%%% Intro paragraph
Quantum computing promises to solve certain classes of problems faster than classical computing \cite{Shor1997, Grover1996}. However, technical imperfections lead to errors that currently prevent most known quantum algorithms from running successfully at scale. Quantum error correction has the potential to allow large codes to run successfully once experiments have surpassed fault tolerance thresholds \cite{Kitaev1997, Aharonov1997, Knill1998, Preskill2006, Xu2023, QuantinuumColorCode}. Of the classes of errors currently preventing scalable fault tolerance, one of the most pernicious is crosstalk, wherein operations (``gates") applied to a target qubit unintentionally also impact one or more additional qubits. These errors are prevalent in many quantum computing platforms\cite{RudingerCrosstalk, DukeCrosstalk, HomeCoherentCancellation, ZZCancelation, Debroy2020, SuperconductingDetuningForCrosstalk}, particularly during parallel gate operation, which is desirable for reducing execution time and correcting qubits with idle errors \cite{ParallelSuperconductingCrosstalk, ParallelMSGates, ParallelVsIdleErrors}. Further, crosstalk errors complicate error correction schemes as they can violate certain assumptions for well-behaved models such as locality and independence of operations \cite{SarovarCrosstalk}. % and Markovianity (? {\color{blue} Definitely does violate that, but not sure it's required for most QEC codes. -MC}) \cite{Debroy2020}.

Previous attempts to reduce crosstalk errors include algorithmic efforts such as crosstalk-aware compiling of circuits \cite{FrequencyAwareCompiling, Debroy2020}, echo-based protocols \cite{MSSpectatorEcho, DukeCrosstalk}, and dynamical decoupling \cite{DynamicalDecouplingCrosstalk}. While these approaches can reduce the impacts of crosstalk, they come at the expense of additional overhead from longer gates and circuits. 
Attempts at physical limitation of crosstalk have relied on either coherent cancellation \cite{QSCOUTManual, ZZCancelation, HomeCoherentCancellation} or highly-engineered independent qubit controls \cite{MEMS-IA, Shih2021, SaffmanInterferingOptics, Shen2013}. These strategies can impose significant experimental burdens to calibrate the system and maintain stability to operate in the low-crosstalk regime.

%thesis
In this work, we describe and implement a physical means of crosstalk mitigation for parallel operation of multi-photon driven quantum gates on a linear chain of trapped ions.
Specifically, the first-order sensitivity to optical crosstalk
% {\color{blue}(OTHER OPTIONS: optical field (spillover), optical crosstalk, electric field amplitude (spillover)), laser electric field spillover}
at neighboring sites is removed 
by choosing distinct single-photon detunings for nearby qubits while maintaining the required two-photon resonance to drive transitions on each target qubit.
% TODO : Paragraph break here?
We demonstrate an order of magnitude improvement in our measured crosstalk when we implement this technique, and we use the improved single-qubit gates with a phase-agnostic two-qubit gate to implement a complete gateset. 
Further, this solution requires no algorithmic overhead nor additional calibrations. 

We implement our solution on individually-addressed ions; however, our result should hold for other platforms using multi-photon transitions with individual qubit addressing, such as neutral atom quantum processors \cite{SaffmanQC}. Analogous ideas for individual addressing of ions have been implemented by applying field gradients to shift each qubit frequency, but these solutions require precise calibrations of each independent qubit frequency and an additional, well-stabilized control field \cite{BGradientCrosstalk, BGradientProposal, ACSGradientProposal, ACSGradientDemonstration, Hensinger2016}. Application of our technique on quantum processors of similar architecture will be immediately useful in mitigating crosstalk without imposing any significant overhead.

%%% Methods
\begin{figure}
    \raggedright
    \begin{tabular}{c c}
        \begin{subfigure}[b]{0.22\textwidth}
        \caption{\raggedright}
        \vspace{0.45cm}
        \centering
        % \raggedright
        \resizebox{1.0\textwidth}{!}{
        \includegraphics[trim=0cm 0.1cm 0cm 0cm]{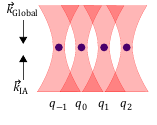}}
        \label{fig:iacartoon}
        \end{subfigure}
    &
        \begin{subfigure}[b]{0.23\textwidth}
        \caption{\raggedright}
        \vspace{-0.4cm}
        \centering
        % \raggedright
        \resizebox{1.0\textwidth}{!}{
        \includegraphics[trim=0cm 0cm 0.4cm 0cm]{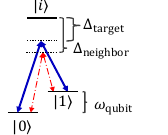}}
        \label{fig:leveldiagram}
        \end{subfigure}
    \end{tabular}
\caption{\raggedright (a) Ions (blue dots) are individually-addressed with tightly-focused laser beams (red shaded regions), which are evenly spaced at 4.5\,$\mu m$. Each beam contains two tones to perform Raman transitions.
A global beam (not shown) illuminates all ions and counter-propagates with the individual-addressing (IA) beams.
Ions are labelled $q_{-1}, q_0, q_1, q_2$ from left to right.
(b) A simplified level diagram shows two-photon Raman transitions between qubit states $\ket{0}$ and $\ket{1}$ through virtual intermediate level $\ket{i}$. To resonantly drive transitions, two frequency components of the applied laser fields (arrows) must have a difference equal to the qubit frequency ($\omega_{\rm qubit}$). When the single-photon detuning for a target ion $\Delta_{\rm target}$ is equal to that of its neighbor $\Delta_{\rm neighbor}$ (not shown), unintended resonant pairs are formed from combinations of the high-intensity control light on the target (solid blue lines) and low-intensity residual illumination from the neighbor (broken red lines). By contrast, when $\Delta_{\rm target} \neq \Delta_{\rm neighbor}$, the only unintended resonant Raman pair is formed from \textit{both} tones of the low-intensity residual illumination.}
\end{figure}

This work is done on the Quantum Scientific Computing Open User Testbed (QSCOUT)\cite{QSCOUTManual}. We use a linear chain of up to four \Yb~ions trapped above a surface-electrode trap. 
Qubits are encoded in the hyperfine ``clock'' ground states\cite{OgYb}. To implement gates, we apply a 355\,nm pulsed laser % with repetition rate (or equivalent frequency comb tooth spacing) $\nu_{\rm rep}/2\pi\approx 120\,$MHz
to drive two-photon Raman transitions\cite{FrequencyCombRaman}. As depicted in Fig.~\ref{fig:iacartoon}, each ion is addressed with a tightly-focused individual addressing (IA) beam. Additionally, a counter-propagating global beam illuminates all ions nearly uniformly (not pictured). Each IA beam and the global beam are independently modulated using their own dedicated acousto-optic modulator (AOM), which converts a radiofrequency (RF) control pulse into a laser pulse. 
More details about the apparatus have been specified in previous work\cite{QSCOUTManual}.

%% Frequency comb discussion (cut)
% To resonantly drive two-photon transitions, the applied laser field must have a frequency component at the qubit frequency, $\omega_{\rm qubit} \approx 2\pi \times 12.6$\, GHz. For single-qubit gates, we achieve two-photon resonance by applying two RF tones ($\omega_0$ and $\omega_1$) to an IA AOM, thus producing two frequency combs (0 and 1) shifted by $\omega_0$ and $\omega_1$. The offset ($\omega_1$ - $\omega_0$) is chosen such that the frequency difference between the $n^{\rm th}$ comb tooth of comb 0 and the $n+105^{\rm th}$ tooth of comb 1 is equal to the qubit frequency: $\omega_1-\omega_0 + 105\nu_{\rm rep} = \omega_{\rm qubit}$. 
To drive two-photon transitions, we drive an AOM with two RF tones ($\omega_0$, $\omega_1$) where the offset (${\omega_1-\omega_0}$) is chosen such that the pulsed laser has a frequency component at the qubit frequency ($\omega_{\rm qubit} \approx 2\pi \times 12.6$\, GHz) \cite{FrequencyCombRaman}.
Applying both tones to an IA AOM produces a co-propagating configuration, which is preferred for single-qubit gates as it is insensitive to the ion's motion. Alternately, we can apply $\omega_0$ to an IA AOM and $\omega_1$ to the global beam AOM in a counter-propagating configuration, which is required for driving the motional sidebands. The effective Raman Rabi rate is proportional to the product of the electric field amplitudes for the two tones: $\Omega_{\rm eff} \propto |E_{\omega_0}||E_{\omega_1}|$. 

To understand the effects of crosstalk during parallel co-propagating single-qubit gates, we consider all resonant Raman pairs that illuminate an ion. 
In typical ion trap experiments, all IA beams are operated with identical single-photon detunings, $\Delta_i = \Delta_j = \Delta$, where $i$ and $j$ are IA beam indices. In this configuration, the leading crosstalk terms come from resonant Raman pairs formed from combinations of the target and neighbor control beams (e.g. $\omega_{0,{\rm target}}$ and $\omega_{1,{\rm neighbor}}$).
These terms have a Rabi rate of order $\epsilon$ where $\epsilon = |E_{i@j}| / |E_{i@i}|$ is the fraction of the electric field amplitude from an IA beam at site $i$ as measured at neighboring site $j$. 
However, as depicted in Fig.~\ref{fig:leveldiagram}, if we operate neighboring IA beams at different single-photon detunings, $\Delta_i \neq \Delta_j$ for $i \neq j$, then each $\omega_{0,i}$ and $\omega_{1,i}$ is distinct from tones applied to neighboring beams and the only resonant Raman pairs form from two tones of the same beam. Therefore, crosstalk is reduced to order $|\epsilon|^2$, as the only unintended resonant pair is formed from residual light of \textit{both} tones from a neighboring beam. 
In other words, uniform single-photon detunings lead to field-sensitive crosstalk while distinct single-photon detunings lead to intensity-sensitive crosstalk.
% In this work, we show that implementing distinct single-photon detunings dramatically reduces crosstalk effects during parallel single-qubit gates. 

\begin{figure}
    \includegraphics[width=0.45\textwidth]{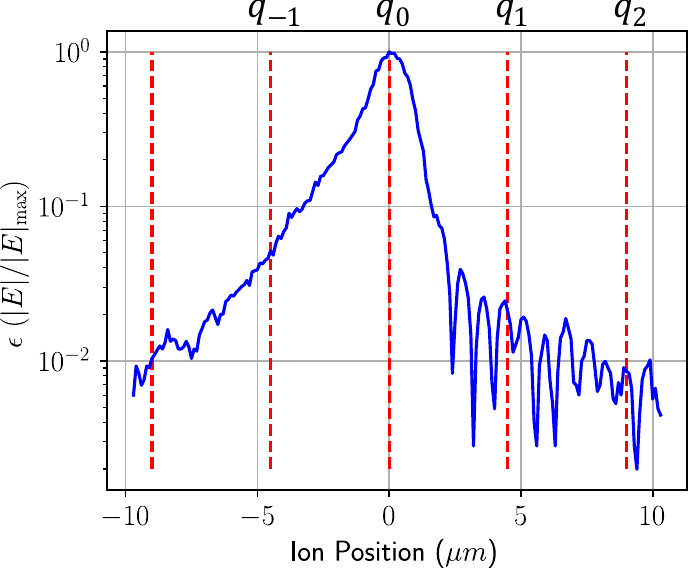}
    \caption{\label{linescan} \raggedright We scan a single ion through a single IA beam (others are off) to estimate the residual illumination at neighboring sites. Beam center locations where ions would typically be located are marked with dashed vertical lines.}
\end{figure}

\begin{figure*}[t]
\centering
\begin{tabular}{c c c c}
    \begin{tabular}[b]{c}
         $\Delta_{i} = \Delta_{j}$ \\
        \begin{minipage}{0.05\textwidth}
        \vspace{1cm}
        \end{minipage}\\
        \resizebox{0.05\textwidth}{!}{
        \includegraphics[trim=0.2cm 0cm 0.2cm 0cm]{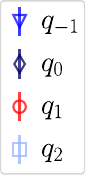}
        } \\
        \begin{minipage}{0.05\textwidth}
        \vspace{1cm}
        \end{minipage}\\
         $\Delta_{i} \neq \Delta_{j}$ \\
         \begin{minipage}{0.05\textwidth}
        \vspace{7.5cm}
        \end{minipage} \\
    \end{tabular}
    &
    %%% Time scan
    \begin{tabular}[b]{c}
    \begin{subfigure}{0.3\textwidth}
        \raggedright
        \caption{\raggedright}
        \vspace{-0.2cm}
        \resizebox{1.0\textwidth}{!}{
        \includegraphics[trim = 0.5cm 0cm 0.5cm 0cm]{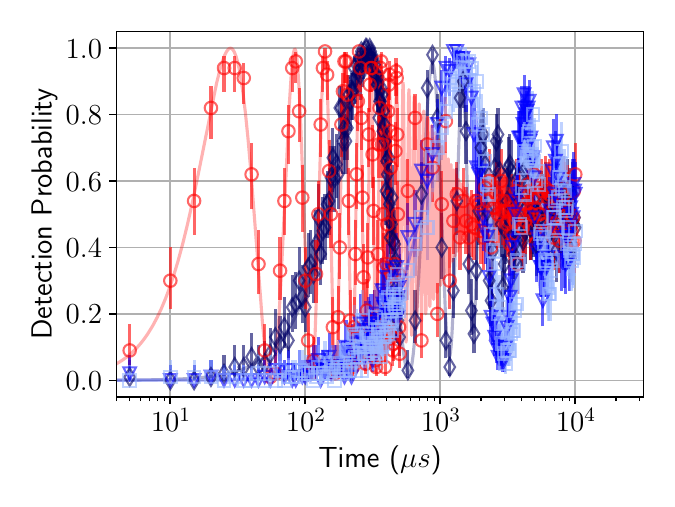}}
        \vspace{-1cm}
        \label{timescan_unmitigated}
    \end{subfigure} \vspace{0.2cm} \\
    \begin{subfigure}{0.3\textwidth}
        \raggedright
        \caption{\raggedright}
        \vspace{-0.2cm}
        \resizebox{1.0\textwidth}{!}{
        \includegraphics[trim = 0.5cm 0cm 0.5cm 0cm]{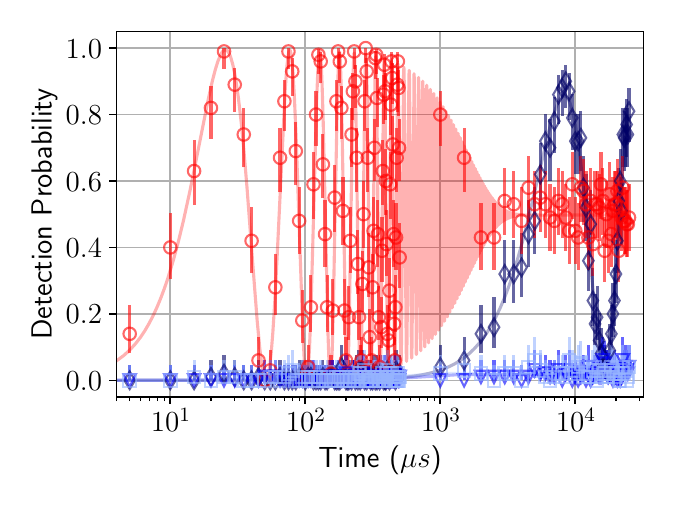}}
        \label{timescan_mitigated}
    \end{subfigure}
    \vspace{1cm}
    \end{tabular}
    & \hspace{-0.5cm}
    %%% Crosstalk matrices
    \begin{tabular}[b]{c}
    \subfloat[$\epsilon$-sensitive configuration]{
        \resizebox{0.25\textwidth}{!}{
        \begin{tabular}{|c|c|c|c|c|}
            \hline
        i & $\Omega_{-1}/\Omega_{i}$ & $\Omega_{0}/\Omega_{i}$ & $\Omega_{1}/\Omega_{i}$ & $\Omega_{2}/\Omega_{i}$ \\
        \hline
        -1 & 1 & 4.45e-03 & 4.07e-03 & 3.58(4)e-03 \\
        \hline
        0 & 7.76e-02 & 1 & 2.04e-02 & 5.20e-03 \\
        \hline
        1 & 1.95e-02 & 9.55e-02 & 1 & 1.77e-02 \\
        \hline
        2 & 1.31e-02 & 3.61e-02 & 1.26e-01 & 1 \\
        \hline
        \end{tabular}
        }
        \label{crosstalkmatrix-unmitigated}
    } \\ 
    \begin{minipage}{0.25\textwidth}
    \vspace{2cm}
    \end{minipage}
    \\
    \subfloat[position=top][$|\epsilon|^2$-sensitive configuration]{ 
        \resizebox{0.25\textwidth}{!}{
    \begin{tabular}{|c|c|c|c|c|}
    \hline
    i & $\Omega_{-1}/\Omega_{i}$ & $\Omega_{0}/\Omega_{i}$ & $\Omega_{1}/\Omega_{i}$ & $\Omega_{2}/\Omega_{i}$ \\
    \hline
    -1 & 1 & 5(5)e-05 & 5(6)e-05 & 1.0(9)e-04 \\
    \hline
    0 & 2.24e-03 & 1 & 1.1(8)e-04 & 1(1)e-04 \\
    \hline
    1 & 5(5)e-05 & 2.55e-03 & 1 & 1(1)e-04 \\
    \hline
    2 & 5(5)e-05 & 8(6)e-05 & 1.98(2)e-03 & 1 \\
    \hline
    \end{tabular}
        }
        \label{crosstalkmatrix-mitigated}
    } \\ 
    \begin{minipage}{0.2\textwidth}
    \vspace{6cm}
    \end{minipage}
    \end{tabular} 
    &
%% Phase scan
    \begin{tabular}[b]{c}
     \begin{subfigure}{0.3\textwidth}
        %\centering
        \raggedright
        \caption{\raggedright}
        \vspace{-0.2cm}
        \resizebox{1.0\textwidth}{!}{
        \includegraphics[trim=0.5cm 0cm 0.5cm 0cm]{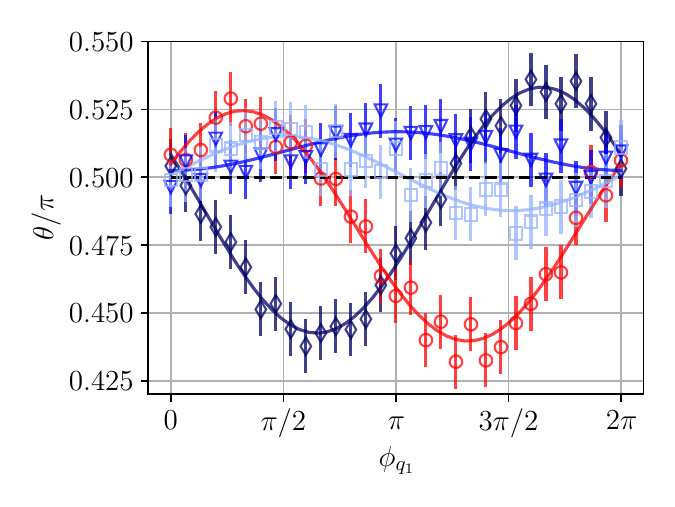}}
        \vspace{-1cm}
        \label{phasescan_unmitigated}
    \end{subfigure} \vspace{0.2cm}
    \\
    \begin{subfigure}{0.3\textwidth}
        %\centering
        \raggedright
        \caption{\raggedright}
        \vspace{-0.2cm}
        \resizebox{1.0\textwidth}{!}{
        \includegraphics[trim=0.5cm 0cm 0.5cm 0cm]{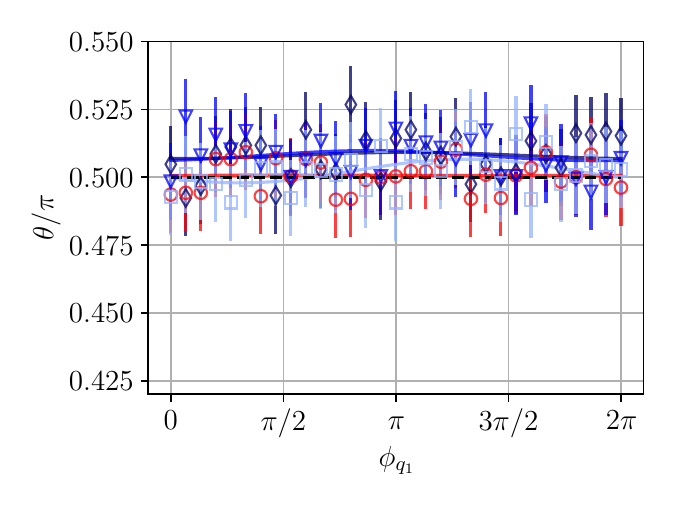}}
        \label{phasescan_mitigated}
    \end{subfigure}
    \vspace{1cm}
    \end{tabular}
\end{tabular}
\vspace{-5.3cm}
\caption{\raggedright Crosstalk effects are reduced during parallel single-qubit gates when run with first-order crosstalk mitigation (bottom, $\Delta_{i}\neq\Delta_j$; b,d,f) compared to without it (top, $\Delta_{i}=\Delta_j$; a,c,e). (a,b) Example Rabi flopping when driving with both tones on $q_1$ (red) and only one tone on all other ions (blue) with crosstalk mitigation (b) compared to without it (a).
(c,d) The crosstalk Rabi rate matrix shows improvement by at least an order of magnitude on all entries in the intensity-sensitive configuration compared to the field-sensitive configuration. Data is collected as in (a,b) by driving qubit $i$ with both tones (and spectators with only one) and measuring the observed Rabi rate, $\Omega_{j}$, on each ion. Unless shown, the fitting uncertainty is $< 1\%$ of each entry. 
(e,f) All qubits are driven with an $R_{\phi}(\frac{\pi}{2})$ pulse in parallel and the Raman phase ($\phi$) is scanned on $q_1$ (red). Without crosstalk mitigation (e), the measured rotation angle ($\theta$) depends strongly on phase
in contrast to when crosstalk is mitigated (f). Curves are fit to a sinusoid to extract rotation error as a function of phase. Horizontal dashed black lines show the ideal $\theta$.
Uncertainty markers on plots are derived from 95\% Wilson score intervals.
}
\end{figure*}

As shown in Fig.~\ref{linescan}, we characterize the spatial extent of an IA beam and estimate the residual illumination at neighboring sites by shuttling a single ion through the beam, measuring the Rabi rate at each point. 
For this measurement, we apply one tone to the IA beam and one tone to the global beam (counter-propagating), such that the Raman Rabi rate is directly proportional to the electric field amplitude of the IA beam, neglecting small variation in the global beam.
From this data, we estimate fractional electric field amplitudes, $\epsilon$, of \LNN{}(\RNN{}) at the left (right) nearest neighbor position, marked $q_{-1}$ ($q_{1}$). 
Asymmetry in the beam profile is caused by optical aberrations.

To implement the intensity-sensitive configuration for co-propagating gates, we shift both tones applied to each IA AOM ($\omega_{0,i}$, $\omega_{1,i}$) such that the difference of each intra-beam pair is constant ($\omega_{1,i}-\omega_{0,i}$=$\omega_{1,j}-\omega_{0,j}$) and each beam resonantly drives Raman transitions for its target ion, but that each $\omega_{0,i}$ and $\omega_{1,i}$ is detuned relative to neighboring beam tones such that no inter-beam resonant pairs are formed. We choose shift frequencies pseudorandomly, sampled over a $\pm0.5$\,MHz range, and check to make sure no nearest neighbor or next nearest neighbors are within 0.1\,MHz. This method is sufficient to satisfy the requirement that the detuning of first-order crosstalk pairs ($|\Delta_i - \Delta_j| \geq$ 0.1\,MHz) is much larger than the crosstalk Rabi rate (order 1\,kHz), while still keeping all drive frequencies close enough to the AOM center frequency to maintain efficiency.
Since the single-photon detuning for the 355\,nm laser is of order 100\,THz, offsets at the MHz level do not significantly alter the target Rabi rate or ac Stark shift.  
We note this solution incurs little experimental overhead and requires no additional calibrations or equipment. 

Next, we directly measure the crosstalk for parallel single-qubit gates in a four-ion chain by applying both tones ($\omega_{0,1}$ and $\omega_{1,1}$) on $q_1$ and a single tone (we use $\omega_{0,i}$ throughout this letter; $\omega_{1,i}$ performs similarly as it has a nearly identical spatial mode) for all spectator ions. We perform this measurement in both field-sensitive and intensity-sensitive configurations, see Figs.~\ref{timescan_unmitigated} and \ref{timescan_mitigated}. 
Without crosstalk mitigation ($\Delta_{i}=\Delta_{j}$), we observe Rabi flopping on the left-nearest neighbor, $q_0$, with $9.6\%$ of the control Rabi rate on $q_1$. By contrast, when we mitigate crosstalk ($\Delta_{i}\neq\Delta_{j}$ for $i\neq j$), the Rabi rate on $q_0$ is reduced to $0.26\%$ of the control Rabi rate on $q_1$. Crosstalk for the right-nearest neighbor ($q_2$) and left, second nearest neighbor ($q_{-1}$) are both reduced from $\approx 2\%$ to $\leq 0.01\%$ of the Rabi rate on $q_1$. 
We repeat this measurement for all ions and record the fractional Rabi rate crosstalk matrix in Figs.~\ref{crosstalkmatrix-unmitigated} and \ref{crosstalkmatrix-mitigated}.

Additionally, we measure the phase dependence of the crosstalk. 
The native single-qubit gate on QSCOUT is a continuously parameterized rotation, $R_{\phi}(\theta)$, as described in reference~\cite{Jaqal}. The rotation axis, $\phi$, is varied by changing the relative phase between the two drive tones
%, $\omega_{0,i}$ and $\omega_{1,i}$ 
(i.e. the ``Raman phase"). 
The rotation angle, $\theta$, is set by the pulse duration. 
In the field-sensitive configuration, identical-frequency target and neighbor light interferes. This interference between the beams depends on the relative optical phase and thus on $\phi$ for each beam. To measure this effect, we apply parallel $R_{\phi}(\frac{\pi}{2})$ gates to all four ions and vary $\phi$ on $q_1$ while fixing $\phi=0$ on all other qubits. The data in Fig.~\ref{phasescan_unmitigated} reveals %not only the optical phase difference between IA beams, but also 
the strong rotation-angle ($\theta$) dependence on phase in the field-sensitive configuration, which causes substantial rotation error for parallel arbitrary-phase gates.
We fit the data for ion $i$ to $\theta_{i}(\phi_{q_1}) = \frac{A_{i}}{2}\sin{(\phi_{q_1}+\xi_{i})}$ where $A_{i}$ and $\xi_{i}$ are free parameters and find the maximum fractional change in rotation angle relative to the average angle ($\bar{\theta}_i$), $(\theta_{i, {\rm max}}-\theta_{i, {\rm min}})/\bar{\theta}_i = A_i/\bar{\theta}_i$. For $\{q_{-1}, q_0, q_1, q_2\}$ we measure $A_i/\bar{\theta}_i =$ \{0.027(5),0.185(4),0.176(6),0.054(4)\}. Consistent with the earlier results, the largest oscillations occur when there is a phase difference between the residual illumination of the phase-varying qubit and its left nearest neighbor as well as when the phase-varying qubit is the left nearest neighbor to another qubit and subject to its residual illumination.
%are on $q_0$ and $q_1$ as varying the phase on $q_1$ dominantly impacts the interference between the $q_0$ beam and $q_1$ residual illumination and that of the $q_1$ beam and $q_2$ residual illumination.
By contrast, in the intensity-sensitive configuration (Fig.~\ref{phasescan_mitigated}), we do not observe strong rotation-angle dependence on phase as all fitted amplitudes are smaller than the 95\% confidence intervals for each point. Fitting results yield  $A_i/\bar{\theta}_i = $\{0.004(8),0.000(6),0.007(7),0.017(7)\}.

For these improvements to be relevant in the context of a working quantum processor, the single-qubit gates also must be compatible with two-qubit gates. 
The native two-qubit gate in QSCOUT is the ubiquitous Mølmer-Sørensen (MS) gate \cite{MS1999}. The MS gate is driven by symmetrically detuned red and blue motional sidebands and requires precise frequency matching conditions of these two drives. Alternately, these conditions can be expressed as precise phase tracking requirements for both sideband drives and their phase relationship to other gates in the circuit. In principle, one could carefully track all of the rotating frames required for the different frequency-shifted single-qubit gates to satisfy these requirements. Instead, we find that use of a phase-agnostic compilation of the MS gate\cite{Lee2005} is sufficient to both combine these single- and two-qubit gates and avoid much of these same field-sensitive crosstalk effects on the two target ions during the MS interaction. 

\begin{figure}
    \raggedright
    \begin{subfigure}[c]{0.42\textwidth}
    %\centering
    \raggedright
    \caption{\raggedright}
    \vspace{-0.2cm}
    \resizebox{1.0\textwidth}{!}{
    \includegraphics{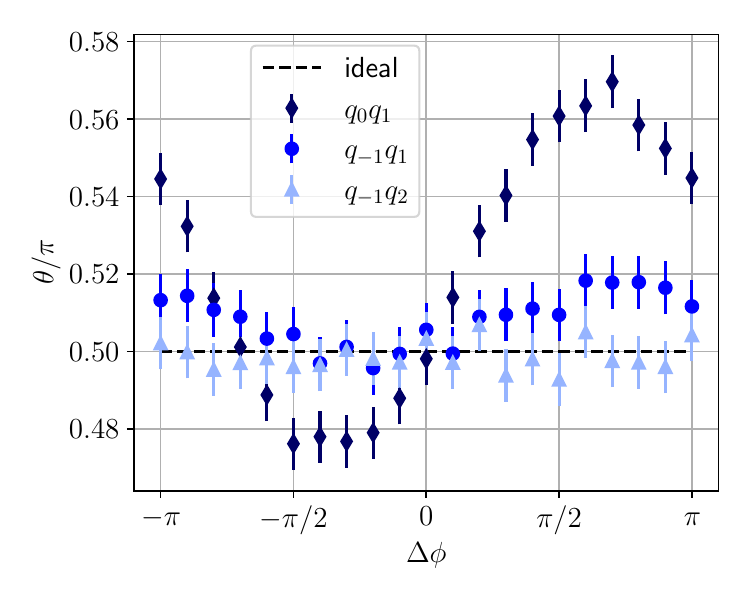}}
    \label{msv}
    \vspace{-0.3cm}
    \end{subfigure}
    \raggedright
    \begin{subfigure}[b]{0.45\textwidth}
    %\centerin
    \raggedright
    \vspace{-0.6cm}
    \caption{\raggedright}
    \resizebox{1.0\textwidth}{!}{
        \Qcircuit @C=.35em @R=.35em {
            & \gate{{R}_{\phi+\frac{\pi}{2}}^{co}(\frac{\pi}{2})} & \gate{{R}_y^{cu}(\frac{\pi}{2})} & \multigate{1}{{MS}_{xx}(\theta)} &  \gate{{R}_y^{cu}(-\frac{\pi}{2})} & \gate{{R}_{\phi+\frac{\pi}{2}}^{co}(-\frac{\pi}{2})} & \qw  \\
            & \gate{{R}_{\phi+\frac{\pi}{2}}^{co}(\frac{\pi}{2})} & \gate{{R}_y^{cu}(\frac{\pi}{2})} & \ghost{{MS}_{xx}^{cu}(\theta)} &  \gate{{R}_y^{cu}(-\frac{\pi}{2})} & \gate{{R}_{\phi+\frac{\pi}{2}}^{co}(-\frac{\pi}{2})} & \qw \gategroup{1}{3}{2}{5}{.4em}{--} \\
            \vspace{3cm} \\
            & & & \zz(\theta) \\ \vspace{8pt}
        }
    }
    \vspace{-0.4cm}
    \label{zzcircuit}
    \vspace{0.3cm}
    \end{subfigure}
    \raggedright
    \begin{subfigure}[c]{0.42\textwidth}
    %\centering
    \raggedright
    \caption{\raggedright}
    \vspace{-0.2cm}
    \resizebox{1.0\textwidth}{!}{
    \includegraphics{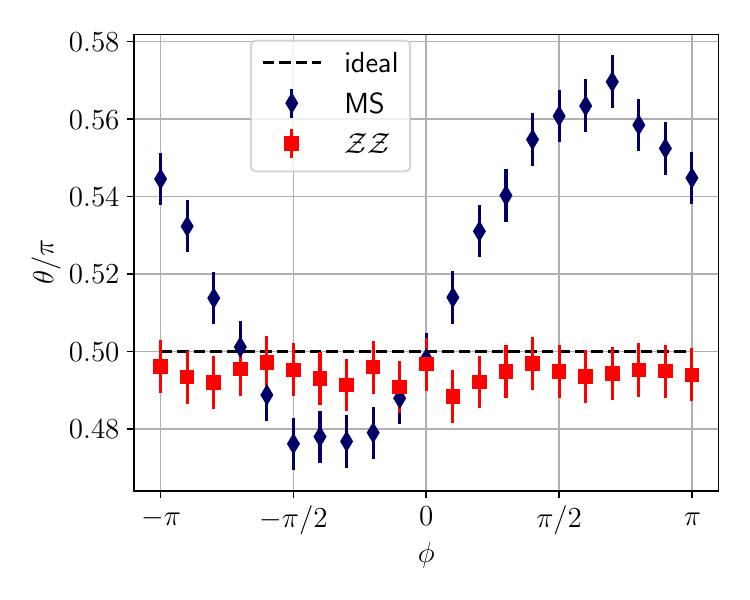}}
    \vspace{-0.7cm}
    
    \label{ms_ams}
    \end{subfigure}
    \raggedright
    
\caption{\raggedright (a) We measure the bare MS gate rotation error for nearest-neighbor, next-nearest-neighbor, and third-nearest-neighbor pairs in a four ion chain. As the target ion separation increases, the crosstalk and resulting phase dependence decreases. 
(b) Circuit diagram for the $\zz(\theta)$ gate. Counter-propagating single-qubit gates ($R^{cu}$) surround the $MS(\theta)$ gate and transform the basis from XX to ZZ. \textit{Optional} co-propagating single-qubit gates ($R^{co}$) surrounding $\zz(\theta)$ transform the gate back to an XX-type operation. 
%and allow for the continuously parameterized $MS(\theta, \phi)$ entangling gate offered on QSCOUT\cite{Jaqal}. 
(c) The measured MS rotation angle on a nearest-neighbor pair ($q_0,q_1$) is constant (squares) when keeping the bare MS interaction at a fixed phase $\Delta\phi$ and applying the rotation-axis phase ($\phi$) to the basis transformation gates, in contrast to the rotation error from phase-dependent crosstalk observed when the phase within the bare MS interaction $\Delta\phi=\phi$ is varied (diamonds). The apparent slight under-rotation of the $\zz$ formulation is likely due to state preparation issues from additional single-qubit infrastructure. Uncertainty markers are derived from 95\% confidence Wilson score intervals.}
\end{figure}

In our current apparatus, all counter-propagating gates must use the global beam, which restricts these gates to a field-sensitive configuration. The MS gate must be counter-propagating to address the motional sidebands, and we implement the gate by turning on the global beam and the two IA beams for the two target ions.
Potentially, one could drive counter-propagating gates with IA beams on both sides and recover second-order crosstalk sensitivity with similar techniques; however, that requires significant apparatus reconfiguration and is outside the scope of this work. Nonetheless, we do characterize the first-order crosstalk sensitivity during the MS interaction in a four-ion chain. 
A dominant error from crosstalk in the MS gate stems from the phase-dependent rotation error on the two target ions of the MS gate, akin to Fig.~\ref{phasescan_unmitigated}. This crosstalk-induced rotation error shrinks as a function of physical distance between the two target qubits, as shown by Fig.~\ref{msv}. 
Other first-order crosstalk effects in the MS gate, such as crosstalk on spectator ions, are left as a subject of future study, but we note that recent demonstration of a spin echo technique to reduce such crosstalk is applicable and compatible with our technique \cite{DukeCrosstalk}. 

Since the QSCOUT system offers the MS gate with a continuous rotation axis ($\phi$) input, the phase-dependent crosstalk on the two target ions can lead to $\phi$-dependent rotation errors if the MS gate is implemented using a phase shift between the two ions ($\Delta\phi$). Similarly, the use of virtual Z gates that advance the phase of all subsequent waveforms\cite{LobserControlHardware} means any MS gates that appear mid-circuit may experience a different phase relationship between the two ions depending on the preceding pulses, regardless of the rotation axis specified. 
% To combat this problem, we compile our codes with a ZZ implementation of the gate, $\zz$, as shown in Fig.~\ref{zzcircuit}\cite{Superstaq}. 
To combat this problem, we use a composite gate, $\zz$, as shown in Fig.~\ref{zzcircuit}, which implements an effective Pauli ZZ interaction. 
We surround the bare MS interaction with counter-propagating (denoted $^{cu}$) single-qubit carrier (qubit transition with no driven motional state change) $R_{\pm y}^{cu}(\frac{\pi}{2})$ pulses which convert the total effective interaction from the XX to ZZ basis\cite{Lee2005,Superstaq}. 
Since ZI and IZ commute with ZZ, prior local phases commute through the $\zz$ gate and therefore do not need to be tracked during the MS interaction. Furthermore, to realize the $\zz$ gate's phase agnosticism, we intentionally ignore the value of local phases tracked by our frame rotations\cite{QSCOUTManual} and virtual Z gates prior to the gate. 
This phase agnosticism allows us to use a fixed relative phase between the two target ions ($\Delta\phi$) at all times.
For circuits requiring an XX-type interaction, a second basis transformation is performed using standard co-propagating single-qubit gates before and after the counter-propagating single-qubit gates, shown in Fig.~\ref{zzcircuit}.
We now calibrate the MS interaction with fixed $\Delta\phi$ and apply the arbitrary phase input ($\phi$) to the basis-transformation single-qubit gates. 
Moving the phase input to the single-qubit gates ideally implements the same unitary as changing the relative phase between the two ions during the bare MS interaction ($\Delta\phi$), but as shown in Fig.~\ref{ms_ams}, the rotation angle is no longer dependent on $\phi$. We note that these basis-transformation gates do not increase the single-qubit overhead as they are also required to combat phase instabilities that arise from path length differences when switching between co-propagating (single-qubit) and counter-propagating (two-qubit) gates\cite{Lee2005}.

\begin{figure}
    \begin{subfigure}[a]{0.48\textwidth}
    \centering
    % \raggedright
    \caption{\raggedright}
    \resizebox{0.8\textwidth}{!}{
    \Qcircuit @C=.35em @R=.35em {
            & \gate{{R}_{y}^{co}(\pi)} & \multigate{1}{{\zz}(\frac{\pi}{2})} &  \gate{{R}_{z}^{co}(\frac{\pi}{2})} &  \gate{{R}_{y}^{co}(-\pi)} & \qw \\
            &  \gate{{R}_{y}^{co}(\frac{\pi}{2})} & \ghost{{\zz}(\frac{\pi}{2})} & \gate{{R}_{y}^{co}(-\frac{\pi}{2})} & \gate{{R}_{x}^{co}(-\frac{\pi}{2})} &\qw
            }
    }
    \vspace{0.2cm}
    
    \label{cnotcircuit}
    \end{subfigure}
    \begin{subfigure}[b]{0.475\textwidth}
    \centering
    \vspace{-0.3cm}
    \caption{\raggedright}
    \vspace{-0.2cm}
    \resizebox{01.0\textwidth}{!}{
    \includegraphics[trim= 1cm 0cm 1cm 0cm]{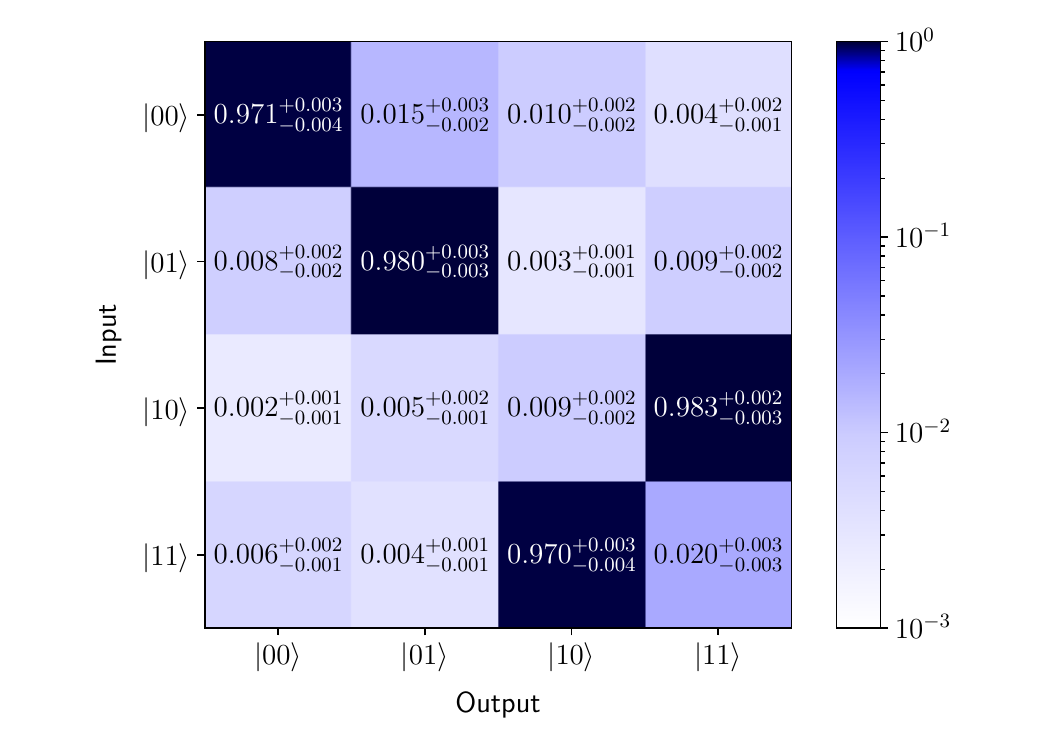}}
    \vspace{-0.75cm}
    
    \label{cnotdata}
    \end{subfigure}
    %\centering
\caption{\raggedright 
(a) Circuit diagram for the composite CNOT gate. We use single-qubit gates and the fully-entangling $\zz(\frac{\pi}{2})$ gate to construct CNOT. 
(b) Population measurements after application of the CNOT gate to each computational basis state demonstrate $\approx$ 96.2\% average fidelity. Uncertainties are 95\% Wilson score intervals. 
}
\end{figure}

To demonstrate that our crosstalk-mitigated single-qubit gates work with the phase-agnostic $\zz$, we run a simple circuit where single-qubit gates convert $\zz$ into a composite CNOT gate as shown in Fig.~\ref{cnotcircuit} \cite{DebnathThesis, Maslov2017}. 
As shown in Fig.~\ref{cnotdata}, we then estimate the fidelity of the CNOT gate from population measurements using the computational-basis states as inputs. 
The 96.2\% average gate fidelity estimate\cite{qutip} is consistent with the fidelity of our bare MS interaction \cite{BalancedGaussian}, indicating that the single-qubit gates are working as expected in concert with the two-qubit gate.

In summary, we demonstrate that the crosstalk on parallel single-qubit gates can be effectively mitigated by driving each individual addressing beam with a distinct single-photon detuning. We observe an order of magnitude or better suppression of all crosstalk Rabi rates as listed in Fig.~\ref{crosstalkmatrix-mitigated}. 
Further, we characterize the rotation error from crosstalk between the two target ions in the arbitrary-phase two-qubit gate and demonstrate that this error can be avoided by applying the phase input to the single-qubit basis transformation gates.
Finally, we demonstrate that the improved single-qubit gates work in concert with our $\zz$ two-qubit gate, forming a universal gateset.
Our technique can be readily adopted on other quantum processors to achieve similar performance gain in parallel gates. This method is also compatible with other algorithmic and physical crosstalk mitigation strategies, allowing for further improvements. As crosstalk is one of many important obstacles on state-of-the-art quantum processors, this work represents a significant step towards achieving scalable fault tolerance.

% As an outlook on the future of this work, we note that similar concepts can be applied by shifting the qubit frequencies with local ac Stark beams. 

%TC:ignore
\begin{acknowledgments}
We thank Rich Rines, Victory Omole, and Pranav Gokhale for discussions inspiring the development of the phase-agnostic MS gates.
We also thank Mallory Harris for helpful discussions in preparation of this work for a general scientific audience.
This research was supported by the U.S. Department of Energy, Office of Science, Office of Advanced Scientific Computing Research Quantum Testbed Program.
% Sandia National Laboratories is a multimission laboratory managed and operated by National Technology \& Engineering Solutions of Sandia, LLC, a wholly owned subsidiary of Honeywell International Inc., for the U.S. Department of Energy's National Nuclear Security Administration under contract DE-NA0003525.  This paper describes objective technical results and analysis. Any subjective views or opinions that might be expressed in the paper do not necessarily represent the views of the U.S. Department of Energy or the United States Government. 
Sandia National Laboratories is a multi-mission laboratory managed and operated by National Technology \& Engineering Solutions of Sandia, LLC (NTESS), a wholly owned subsidiary of Honeywell International Inc., for the U.S. Department of Energy’s National Nuclear Security Administration (DOE/NNSA) under contract DE-NA0003525. This written work is authored by an employee of NTESS. The employee, not NTESS, owns the right, title and interest in and to the written work and is responsible for its contents. Any subjective views or opinions that might be expressed in the written work do not necessarily represent the views of the U.S. Government. The publisher acknowledges that the U.S. Government retains a non-exclusive, paid-up, irrevocable, world-wide license to publish or reproduce the published form of this written work or allow others to do so, for U.S. Government purposes. The DOE will provide public access to results of federally sponsored research in accordance with the DOE Public Access Plan.
SAND2023-09857O
\end{acknowledgments}

\section*{Author Declarations}
\subsection*{Conflict of Interest}
The authors have no conflicts to disclose.

\subsection*{Author Contributions}
\textbf{Matthew N. H. Chow:} Conceptualization (lead); Data Curation (equal); Formal Analysis (equal); Investigation (equal); Methodology (equal); Software (equal); Validation (equal); Visualization (equal); Writing - original draft (lead); Writing - review \& editing (equal).  
\textbf{Christopher G. Yale:} Conceptualization (equal); Data Curation (equal); Formal Analysis (equal); Investigation (equal); Methodology (equal); Project Administration (supporting); Software (equal); Supervision (supporting); Validation (equal); Visualization (lead); Writing - original draft (supporting); Writing - review \& editing (equal).
\textbf{Ashlyn D. Burch:} Conceptualization (supporting); Data Curation (equal); Investigation (equal); Methodology (equal); Software (equal); Validation (equal); Visualization (supporting); Writing - review \& editing (equal).
\textbf{Megan Ivory:} Validation (supporting); Writing - review \& editing (equal).
\textbf{Daniel S. Lobser:} Conceptualization (equal); Methodology (supporting); Software (lead); Validation (equal); Writing - review \& editing (equal).
\textbf{Melissa C. Revelle:} Conceptualization (supporting); Methodology (supporting); Validation (supporting); Visualization (supporting); Writing - review \& editing (equal).
\textbf{Susan M. Clark:} Conceptualization (supporting); Funding Acquisition; Investigation (equal); Methodology (supporting); Project Administration (lead); Software (supporting); Supervision (lead); Validation (equal); Visualization (supporting); Writing - review \& editing (equal).  

\section*{Data Availability Statement}
The data that support the findings of this study are available from the corresponding author upon reasonable request.

\bibliography{main}
% \detailtexcount{main}
%TC:endignore

\end{document}